\documentstyle[psfig]{mn}

\newcommand{\ale}{\ \raisebox{-.3ex}{$\stackrel{<}{\scriptstyle \sim}$}\ }

\title[Magnetically warped discs]{Magnetically warped discs in close binaries}
	
\author[James R. Murray, Dalia Chakrabarty, Graham A. Wynn \& Louisa Kramer ]
{James R. Murray\thanks{email address: jmu@star.le.ac.uk}, Dalia
Chakrabarty, Graham A. Wynn
 \& Louisa Kramer\\
University of Leicester,University Road, Leicester LE1 7RH, UK\\}	
\begin{document}

\maketitle

\begin{abstract} 
We demonstrate that measurable vertical structure can be excited in the
accretion disc of a close binary system
by a dipolar magnetic field centred on the secondary
star. We present the first high resolution hydrodynamic simulations  
to show the initial development of a uniform warp in a tidally
truncated accretion disc. The warp
precesses retrogradely with respect to the inertial frame. The
amplitude  depends on the phase of the warp with respect to the binary frame. A
warped disc is the best available explanation for negative superhumps.

\end{abstract}


\section{Introduction}

Much of the variability of cataclysmic variables can be traced to
instabilities in the accretion disc. Thermal-viscous instability leads to
dwarf nova outbursts, and tidal instability leads to superhumps and
superoutbursts. More recently, people have been giving thought to
warp-type instabilities. Pringle (1996) showed that radiation from the
central accreting object will exert an asymmetric force on either
surface of the disc that can be
sufficent to warp it. However, relatively extreme conditions are
required to generate a warp in this way (Ogilvie \& Dubus 2001). 
In this paper we show numerically that stellar magnetic
fields can  warp the disc of a cataclysmic variable. In particular we
shall focus on the effects of the secondary (donor) star's magnetic field.

The light-curves of many cataclysmic variables exhibit periodicities
close to but distinct from the orbital period. These signals are
categorised as {\em positive superhumps} if they are longer than
orbital and {\em negative superhumps} if they are less than
orbital. Both are thought to emanate from the disc with the well
established explanation for the former being that they are the visible
consequences of an eccentric disc precessing with respect to the binary's
tidal field (see O'Donoghue 2000 for a review). 
Negative superhumps have not been studied in nearly as
much detail.
The observations of negative superhumps are summarised in Patterson
(1999) in which 11 systems are listed. With periods typically a few
per cent less than orbital, a key feature of negative
superhumps is that they appear to be independent of positive
superhumps. They have been observed both in the absence of and in
conjunction with their better studied brethren.  

Patterson et al. (1993) suggested that negative superhumps occur when
the accretion disc is tilted to the binary plane. Just as the line of
nodes of the Moon's orbit about the Earth regresses, so too
the intersection of the disc and binary planes precesses
retrogradely. The effect is present in the purely dynamical problem 
and is analogous to the eccentric disc case. A test particle in circular
orbit around a single star has three natural frequencies, that of its
azimuthal motion ($\omega$), and those of any radial or vertical
motion. Around a single star these frequencies are identical and any
perturbation to a circular orbit causes the particle to orbit in an
ellipse. For a particle in orbit around a star in a binary 
we find that the tidal field breaks the symmetry of the three
periods. The radial frequency becomes
slightly lower than ($\omega$) and the vertical frequency slightly
higher. As a result, the axis of any perturbation to the orbit will
precess. Analytical calculations (e.g. Papaloizou \& Terquem 1995) and
numerical simulations (e.g. Larwood et al. 1996; Wood et al. 2000)
showed that an initially rigidly tilted disc in a binary system will
precess retrogradely, approximately as a rigid body, as a result of
the gravitational torque acting on the disc.

The major stumbling block for the warped disc hypothesis has been the
lack of a viable mechanism for generating it. Three possibilities can be explored. A warp
might have developed as a result of an impulsive torque or out of the
plane initial condition, and then precessed freely (see e.g. Larwood
et al. 1996). A second possibility is that the warp developed from a 
co-planar disc via an instability. Lubow (1992) and Pringle (1996)
discussed dynamical and radiative disc instabilities respectively. 
The third possibility, the one explored in this paper, is that
the warp is excited by continual forcing out of the plane.

Both  impulsively and resonantly generated warps would be forced
back into the plane as the disc evolved viscously. Murray \& Armitage
(1998) presented a series of calculations suggesting that the dynamical and
radiative instabilities were too weak to overcome viscosity. Whilst
continual forcing could overcome the problem of viscous flattening,
the long term disc response will not be free precession but be with
the same period as the forcing.

The inspiration for this paper was the idea espoused by Terquem and
Papaloizou (2000), that a warp might be induced in a circumstellar
disc by the stellar magnetic field. Should the axis of the magnetic
field be misaligned with the binary rotation axis, then the magnetic
field will exert a torque in the plane of the accretion disc.  
Terquem and
Papaloizou were interested in  T Tauri stars, however the principle
could equally be applied to a cataclysmic variable with a moderately
magnetised accreting white dwarf. Lai (1999) described in detail the
mechanisms by which the stellar magnetic field would couple with the
disc. Again however his attention was directed more at X-ray accreting
pulsars and not at cataclysmic binaries.

We carried out an initial series of smoothed particle
hydrodynamics simulations and did indeed find that vertical structure
was induced in the disc. However this structure 
co-rotated with the accreting star and could not possibly generate the
periods characteristic of negative superhumps. We will describe these
simulations in a subsequent paper (Wynn, Murray \& Chakrabarty in preparation). 
Pringle (private
communication) suggested we allow the donor rather than the accretor
to be magnetised.
Whereas the primary star in a cataclysmic variable typically rotates several
times faster than the binary, the secondary is tidally locked and
hence co-rotates with the binary. This paper describes a series of
simulations with which we sought to determine whether the resultant
accretion disc structure was {\em approximately} stationary in 
the binary or in the inertial frame. 

Several authors have already performed detailed analysis of the
suitability of the smoothed particle hydrodynamics (SPH) technique for
modelling three dimensional structure in accretion discs. Murray
(1997) and Yukawa, Boffin \& Matsuda (1997) demonstrated that tidally
excited spiral structure is resolved in well constructed two
dimensional and three dimensional SPH
simulations respectively.
Larwood et al. (1996) modelled the precession of tilted discs in close
binaries. They  followed the long term evolution of discs tilted at 45
and 90 degrees to the binary plane using $17500$ particle simulations.
Nelson \& Papaloizou (1999) studied the propagation of
bending waves in accretion discs. They specifically tested SPH
simulations against the linear theory of warped discs and found good
agreement. Calculation resolution ranged from $20000$ to $102000$
particles. They later (Nelson \& Papaloizou 2000) studied the warp
produced by the Bardeen-Petterson effect using simulations ranging up
to $200000$ particles in size. Wood, Montgomery \& Simpson (2000)
used simulations with 25000 particle resolution to follow the
precession of a pre-tilted disc in a close binary.

The calculations described later in this paper involve over
$500000$ particles. At this resolution we are not limited to resolving
the gross features of the warp itself. We can also determine the 
 the vertical structure caused by the tidal forces of the secondary, 
and then follow the interaction between the two.

\section{Magnetised secondaries}

The secondary stars in cataclysmic variables have rotational
periods which are tidally locked to the binary's orbital 
period, and are typically of the order of a few hours. 
The connection between rapid stellar rotation rates and 
chromospheric, and by implication magnetic, activity has long been 
established (eg. Kraft, 1967; Skumanich, 1972). This connection is often 
invoked as an argument in favour of the magnetic nature of the 
rapidly rotating secondary stars in CVs. 
Direct observational evidence for magnetic
activity remains elusive however. Some of the stongest evidence
comes from observations of star spots (and their close association
with magnetic activity) in the CV ST Leonis Minoris (Howell et al.,
2000)  and the pre-CV V471 Tau (Applegate \& Patterson, 1987) 
as well as in a number of other types of close binary systems. 
The evidence for solar-type activity cycles in the secondary
stars of CVs is discussed by Bianchini (1990). Any such cycle would 
imply the presence  of significant magnetic fields. The existance
of these magnetic fields has strong theoretical support.
Angular momentum loss by magnetic braking of the secondary star
is the generally accepted mechanism driving the orbital
evolution of CVs above the orbital period gap (Verbunt \& 
Zwann, 1981; Mestel \& Spruit, 1987). Moreover, theories
which attempt to explain the sychronism between the binary orbital 
period and the rotational periods of magnetic white dwarfs in CVs 
require secondary magnetic moments of the order 
$\mu_2 \sim 10^{33}$ - $10^{34}$ G cm$^3$ (eg. King, Whitehurst \&
 Frank, 1990). 

For the purposes of the simulations presented in this paper we assume
that the secondary stars in CVs have a significant dipolar magnetic
field, and that the magnetic moment is similar to that quoted above.
This implies that the (unperturbed) field strengths close to the edge 
of the accretion disc are of order $\sim 10^2$ G. As the gas in the 
accretion disc orbits the white dwarf in the binary it will twist the
field of the secondary star and induce currents within the disc. The
interaction of these currents and the imposed magnetic field gives 
rise to a variety of forces. King (1993) developed a drag
prescription to parametrise the effects of these magnetic forces on
accreting gas in the AM Herculis sub-class of cataclysmic
variables. 
Pearson, Wynn and 
King (1997) later made use of this prescription to study the effects
of the azimuthal force arising from
the interaction of radial surface currents and the vertical component 
of the magnetic field on the structure of a two dimensional disc in the 
orbital plane of the binary. This study showed that disc-field
interaction caused the disc to lose angular momentum (to the 
secondary star via the field lines) causing the disc radius
to decrease, and enhancing the mass accretion rate. Moreover,
resonance phenomena similar to those in the SU~UMa systems
caused the disc to become eccentric. In this paper we extend the 
analysis of the effect of the magentic field on the accretion disc
to three dimensions. 

Lai (1999) made a detailed analytic study of the torques to which
a disc is subjected to when an external stellar field is imposed
upon it. In the general case he found that a warping torque resulted 
from the interaction of the azimuthal component of the stellar 
magnetic field and the radial surface current on the disc induced by 
the twisting of the threaded vertical field component. However, the
details of the disc-field interaction remain unclear, in particular 
there is little agreement on the range and strength of the magnetic 
stresses over the disc. In order to examine the response of a fluid 
disc to an applied warping torque we implemented a simplified version 
of the magnetic interaction in which the gas within the disc is
subject to an acceleration of the general form
\begin{equation}
{\rm {\textbf a}_{mag}} \simeq -k [\textbf{v}-\textbf{v}_{f}]_{\bot}
\label{aeqn}
\end{equation}
where $\textbf{v}$ and  $\rm \textbf{v}_{f}$ are the velocities
of the material and field lines respectively, and the suffix $\bot$ refers to the 
velocity components perpendicular to the field lines. The parameter
$k \sim 1/t_{\rm mag}$ reflects the details of the plasma-magnetic 
field interaction, and for the purposes of this paper 
we adopt the simple scaling $k \propto B^2(r,\theta,\phi)$ and
assume that the magnetic field has a purely dipolar form.
We can estimate the magnitude of $k$ in the vicinity of the 
accretion disc in a cataclysmic variable with a magnetic secondary
star from the comparison of (\ref{aeqn}) and the calculations
of Lai (1999). Equating the force per unit area in each case
we have
\begin{equation}
\Sigma k [\textbf{v}-\textbf{v}_{f}]_{\bot} \sim \frac{\mu_2^2}
{2\pi r_2^6}
\label{mageq} 
\end{equation}
where $r_2$ is the distance from the secondary star and $\Sigma$ is
the surface density of the accretion disc. Typically $r_2 \sim 10^{10}$
cm, $\mu_2 \sim 10^{34}$ G cm$^3$, $\Sigma \sim 100$ g cm$^{-2}$ and
$|[\textbf{v}-\textbf{v}_{f}]_{\bot}| \sim 100$ km s$^{-1}$ in a
CV giving $k(10^{10}\: {\rm cm}) \sim 10^{-2}$ s$^{-1}$. This simplistic form of the 
magnetic force preserves all of the components required to induce 
a warp within the disc and has a magnitude of the order of that
expected of magnetic stresses induced on the accretion disc by 
a magnetic secondary in a CV.

\section{Code aspects}
We use a smoothed particle hydrodynamics code that has been described
extensively elsewhere (Truss et al. 2000, Murray 1996). The code uses
operator splitting to separate the calculation of the gravitational
and magnetic forces from that of the gas-dynamical processes. The code
has been parallelised using OpenMP to run on shared memory 
parallel supercomputers. The viscosity is slightly modified from
Murray (1996). We are free to choose the length scale in the viscosity
term so we set it to equal the disc scale height, and so are able to
reproduce a Shakura-Sunyaev $\alpha$ viscosity. The
viscosity term also produces a bulk viscosity of similar magnitude in
regions where the velocity field is significantly divergent. 
A variable smoothing length $h$ is
used. To ensure the disc is everywhere resolved $h$ must be
less than the scale height whilst maintaining an adequate number of
neighbours.

We use an isothermal equation of state. In other words we assume any
viscously dissipated energy is instantly radiated away.

The magnetic interaction is implemented by including the acceleration
given by equation~\ref{mageq} along with the gravitational terms in
the operator split equations of motion.

As in previous papers we have normalised units so that the binary
separation $d$, the total system mass $M$, and the binary angular
velocity $\Omega_b$ are all equal to one.

\section{Calculations}

\begin{figure}
\psfig{file=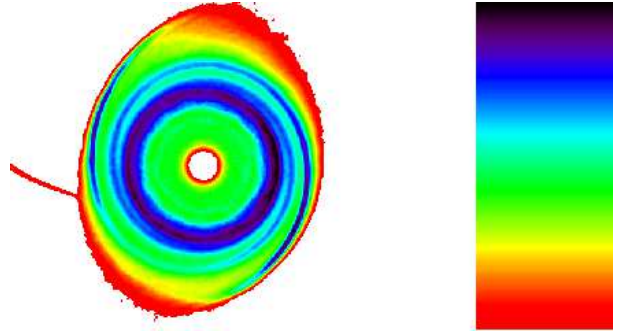,width=8cm}
\caption{Surface density map of the initial state. 
Note the finely
resolved tidal spiral arms. There are 536033 particles in the
disc. Midway into the disc there is a ring where the surface density
is a maximum. This is an artefact from the disc's creation. 
The key to the right shows the colour coding from low
density (red) to high density (black).}
\label{disc1}
\end{figure}

\begin{figure}
\psfig{file=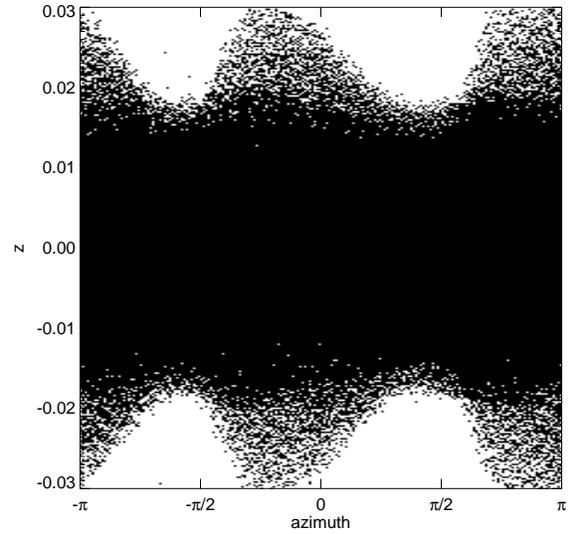,width=8cm}
\caption{Silhouette of the planar (unwarped) disc as seen by an observer in the
binary plane. Azimuth
$-\pi$ is along the binary axis between the stars, and then increases 
anticlockwise round the disc.}
\label{nonwarp}
\end{figure}

\begin{figure*}
\psfig{file=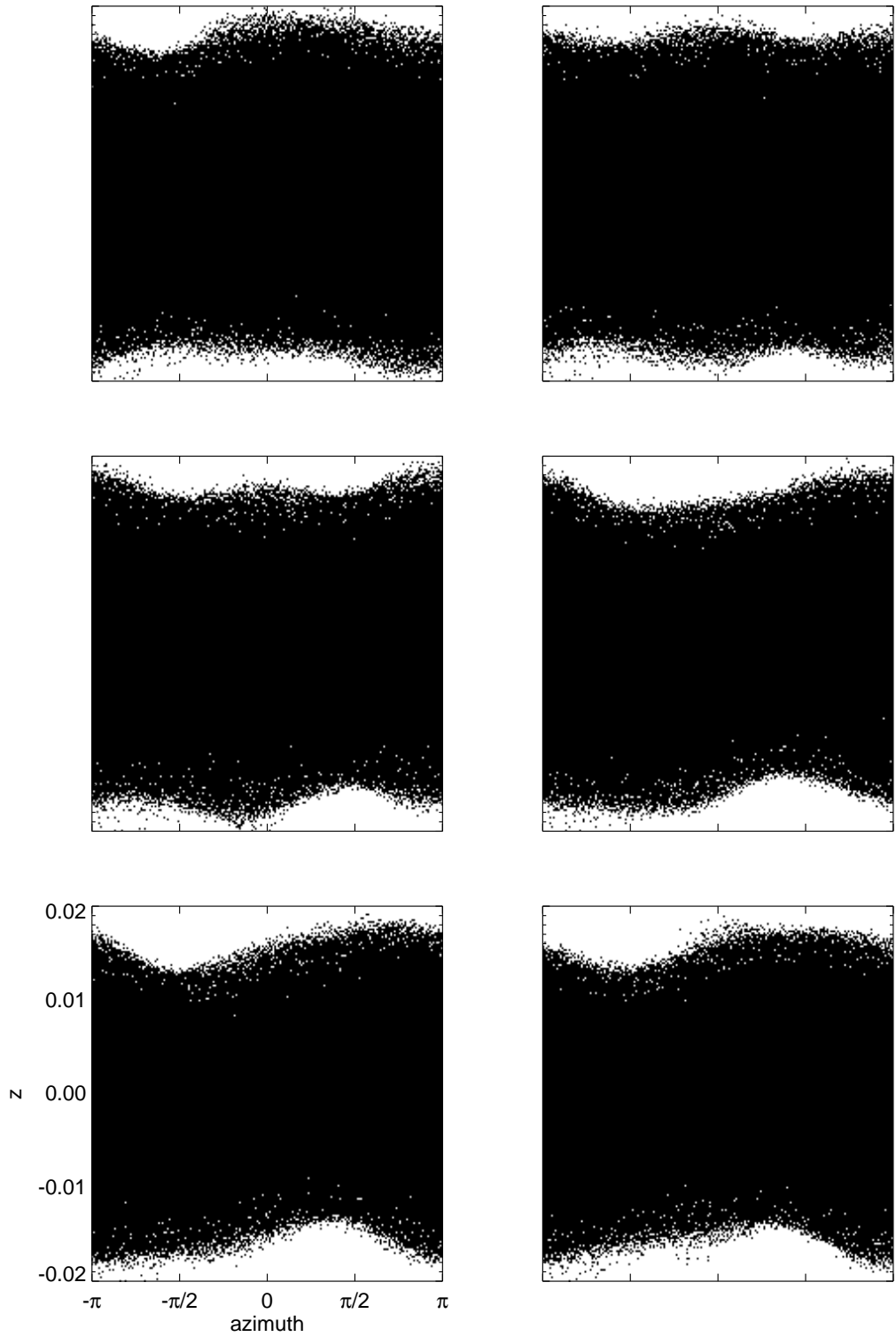,width=15cm}
\caption{Disc silhouette at times t=5,6,7,8,9 and 10 (from top left to
bottom right). Azimuth is measured as in figure~\ref{nonwarp}.}
\label{sequence}
\end{figure*}

\begin{figure*}
\psfig{file=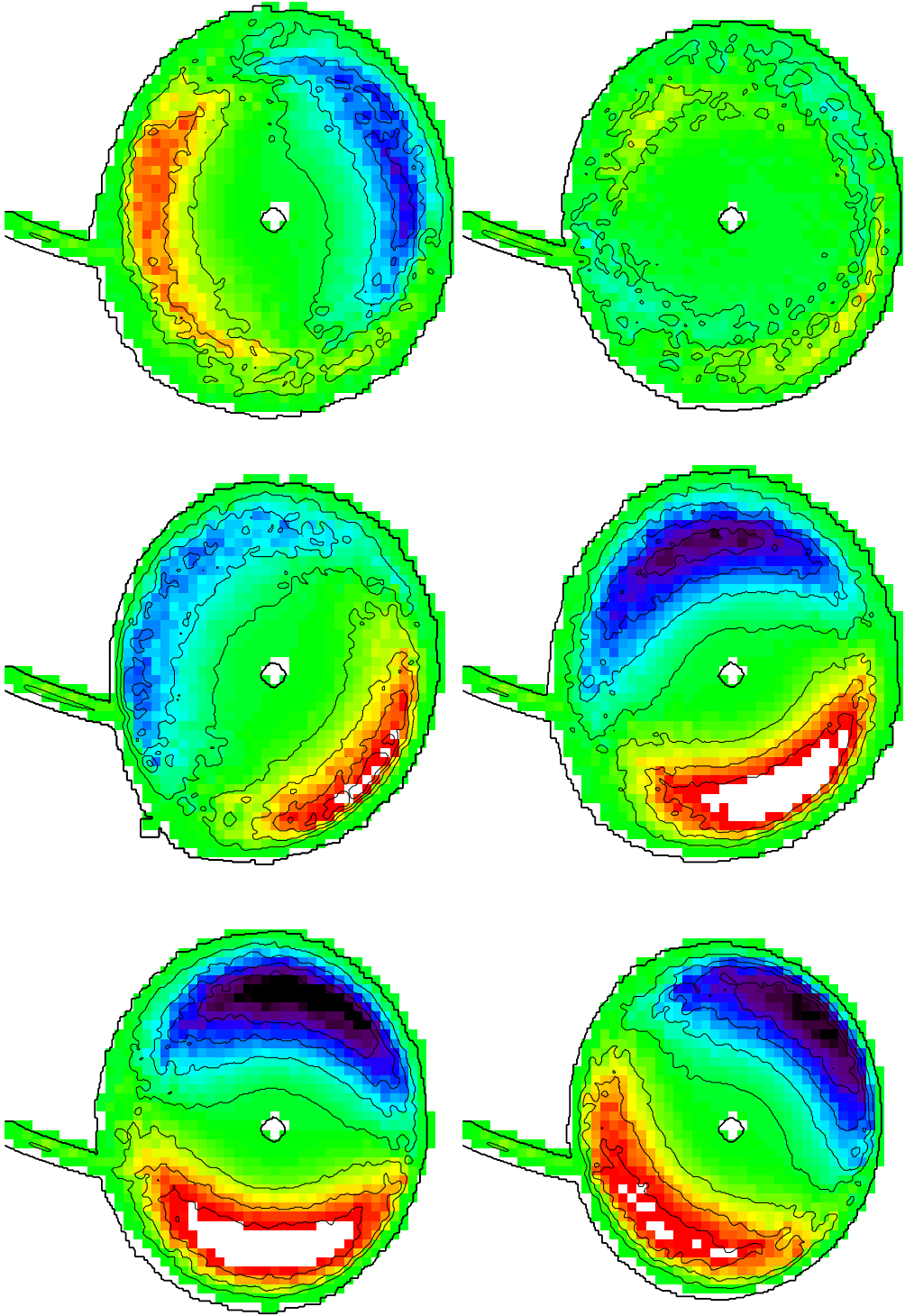,width=15cm}
\caption{ Vertical structure of the disc at times t=5,6,7,8,9 and 10 (from top left to
bottom right).  Blue hues indicate structure below the disc
plane, green shows regions that are symmetric about $z=0$ whilst
yellow and red show regions that are progressively more lifted above
the disc plane. The panels in this figure correspond to those in 
figure~\ref{sequence}.}
\label{height1}
\end{figure*}
 
The system and code parameters for the calculations are listed 
in Table~\ref{tab:params}. 
We constructed a three dimensional initial state for the calculation
by running a two dimensional non-magnetic simulation to
equilibrium. The particles were then replicated in z to produce a
three dimensional disc which was then followed for several orbits to
allow vertical oscillations to die out. The resulting disc comprised 
536033 particles. It is shown in plan view, shaded according to
surface density, in figure~\ref{disc1}.
Tidally inspired spiral shocks are clearly visible. We see no evidence
that spiral shocks are suppressed or any weaker in three dimensions
than in two dimensional calculations.

We also show the silhouette of the  initial disc
(figure~\ref{nonwarp}) as would be seen by an
observer on the disc plane. This figure was obtained by plotting all
the particles on the $\theta-z$ plane. The two spiral arms produced
50 per cent increases in the disc height which should be observable in
X-ray and EUV observations of eclipsing systems (see e.g. Billington
et al., 1996).

As we shall see below, the starting disc
was not in equilibrium with the mass addition stream from the secondary
star. However the time-scale for mass evolution is much longer than
that for magnetically induced evolution and this did not affect our results.

We completed two simulations; a reference calculation with no magnetic
field, and a second in which the magnetic field was added
instantaneously.  
The dipole orientation
\begin{equation}
 \hat{\bf \mu } = \sin (\phi_{\rm d}) (\cos (\theta_0 + \Omega_b t)\,{\bf i}
+ \sin (\theta_0 + \Omega_b t)\,{\bf j}) + \cos (\phi_{\rm d})\,{\bf k}
\end{equation}
where  the angular speed of the binary $\Omega_b=1$. 
${\bf i}$,${\bf j}$ and ${\bf k}$ are the cartesian unit vectors. In
our coordinates the disc lies in the x-y plane with the angular
momentum vector of the binary in the positive z direction. 

Mass transfer from the secondary was incorporated into the simulations by
adding single particles from $L_1$ at intervals
$\Delta\,t=0.01\,P_{\rm orb}$. 
As in previous simulations
(e.g. Murray, 1996), the initial velocity of injected particles was
taken from Lubow \& Shu (1977). The disc was not in equilibrium with
the mass transfer stream which was only incorporated into the
simulation as a useful tracer of the secondary magnetic field.
Gas flow on the secondary star was not
resolved in the calculations.

The
calculations described here (and their antecedents) took the order 
of one week of wall time on 8
processors of an SGI Origin 2000 or 3000.

\begin{table}
\caption{Simulation parameters. The magnetic dipole on the secondary is
oriented so as to be
$\phi_{\rm d}$ to the vertical and $\theta_{\rm d}$
prograde of the binary axis.  }
\label{tab:params}
\begin{tabular}{ccc}
parameter & symbol &value\\
\hline
mass ratio & $q$ & 0.4 \\
sound speed & $c$ & 0.05 $d\,\Omega_b$\\
viscosity parameter & $\alpha$ & 1.0 \\
magnetic parameter  & $k$ & 10$^{-5}$ (10$^{-2}$ s$^{-1}$)\\
dipole declination &$\phi_{\rm d}$ & $20 \deg$ \\
dipole azimuth &$\theta_{\rm d}$ & $15 \deg$ \\
\hline
\end{tabular}
\end{table}

\section{Results}

\begin{figure}
\psfig{file=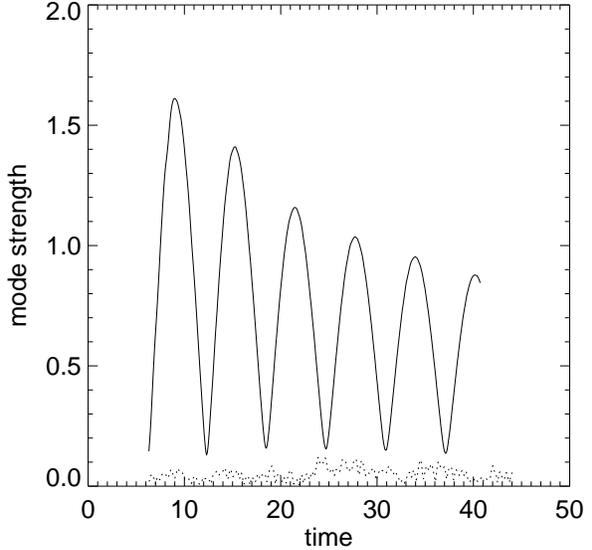,width=8cm}
\caption{Strength of the warp as a function of time when the secondary
is magnetised (solid line) and unmagnetised (dotted line). Both curves
have been normalised by multiplying by $10^3$.}
\label{discwarp}
\end{figure}

\begin{figure}
\psfig{file=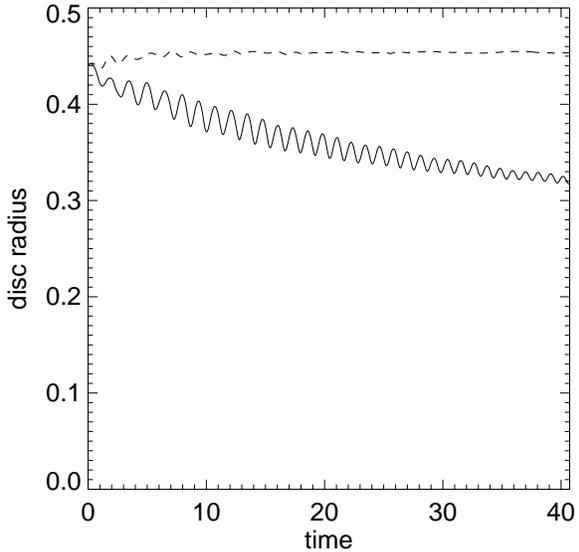,width=8cm}
\caption{Average disc radius for the magnetised (solid line) and unmagnetised
(dotted line) simulations.}
\label{discrad}
\end{figure}


In this section we show that the drag on an accretion disc, exerted
by a tilted magnetic dipole anchored to the secondary star, results in a warp.
Figure~\ref{sequence} is a sequence of six disc silhouettes, each
generated in the same manner as figure~\ref{nonwarp}. The
snapshots show the disc at equally spaced times from $t=5-10$ 
(in our units $P_{\rm orb}=2\pi$). Whilst the disc is approximately
symmetric about its midplane in the first three panels, by the fourth ($t=8$)
it is warped. 

We have plotted six plan
views of the disc vertical structure in figure~\ref{height1}, 
corresponding to the snapshots in figure~\ref{sequence}. Each
image is constructed by interpolating particle $z$ (averaged over all
particles in a region) to a grid.  Blue
regions of the disc are predominantly below the $z=0$ plane. Green
regions are approximately symmetric about the midplane, and yellow-red
regions are more extended above $z=0$. The magnetic torque deflects
the stream of gas from the secondary above the plane (redwards).
Contours have also been drawn to aid
the eye.  Figures~\ref{sequence} and~\ref{height1} therefore show the disc's
initial response to an instantaneously applied out of plane force.

Less than one binary period after the magnetic 
force was applied, the disc warp was a coherent global structure (panel
1). However one sixth of a binary period later (panel 2) the warp was no longer apparent.
It reappeared again, much stronger, in panels three and four.
In each image the warp is a disc-wide structure, with the spiral density
waves in the initial disc having been rapidly diminished by the warp.
 
If one follows, for example, 
the yellow-red zone
from panel to panel one sees that the warp has precessed
retrogradely {\em in the binary frame}. The disc precession with respect to
the inertial frame occurred on a much longer time scale so
quantitative analysis of the calculations was required.

We Fourier
transformed the disc's vertical structure in both azimuth
and time, as in Murray \& Armitage (1998). 

The sine transform of the vertical displacement,
\begin{equation}
S^z_{\sin}(t)=\frac{1}{\pi 
N }
\int_t^{t+2\pi} (\sum_{p=1}^N z_p(t')\sin(\theta_p)) \,\rm{dt'},
\label{eq:warp}
\end{equation}
where $\theta_p$ is the angular position of particle $p$ (measured in
the inertial frame) and $N$ is
the total number of particles. The cosine transform is defined
similarly and the mode strength is set equal to the root mean square
of the two. The mode strength has units of length, and so a large
response would be of the order of the disc height ($0.01\,d$). As a
check, we returned to the initial disc
(as shown in figure~\ref{disc1}) and applied a uniform tilt of 
five degrees to it. We then evolved the tilted disc for one orbital
period, and applied equation~\ref{eq:warp}. We found $S^z \simeq 0.02$ which
agreed with the average vertical displacement of the particles.

Figure~\ref{discwarp} overlays the warp mode strength for
the two calculations. For convenience only we have normalised both
curves, by a factor $10^3$. No warp developed in the reference calculation,
and the corresponding curve (dashed line) is noisy  but with
typical magnitude  of only $10^{-4}$ or a factor two hundred smaller than for
the five degree warp.  

\begin{figure*}
\psfig{file=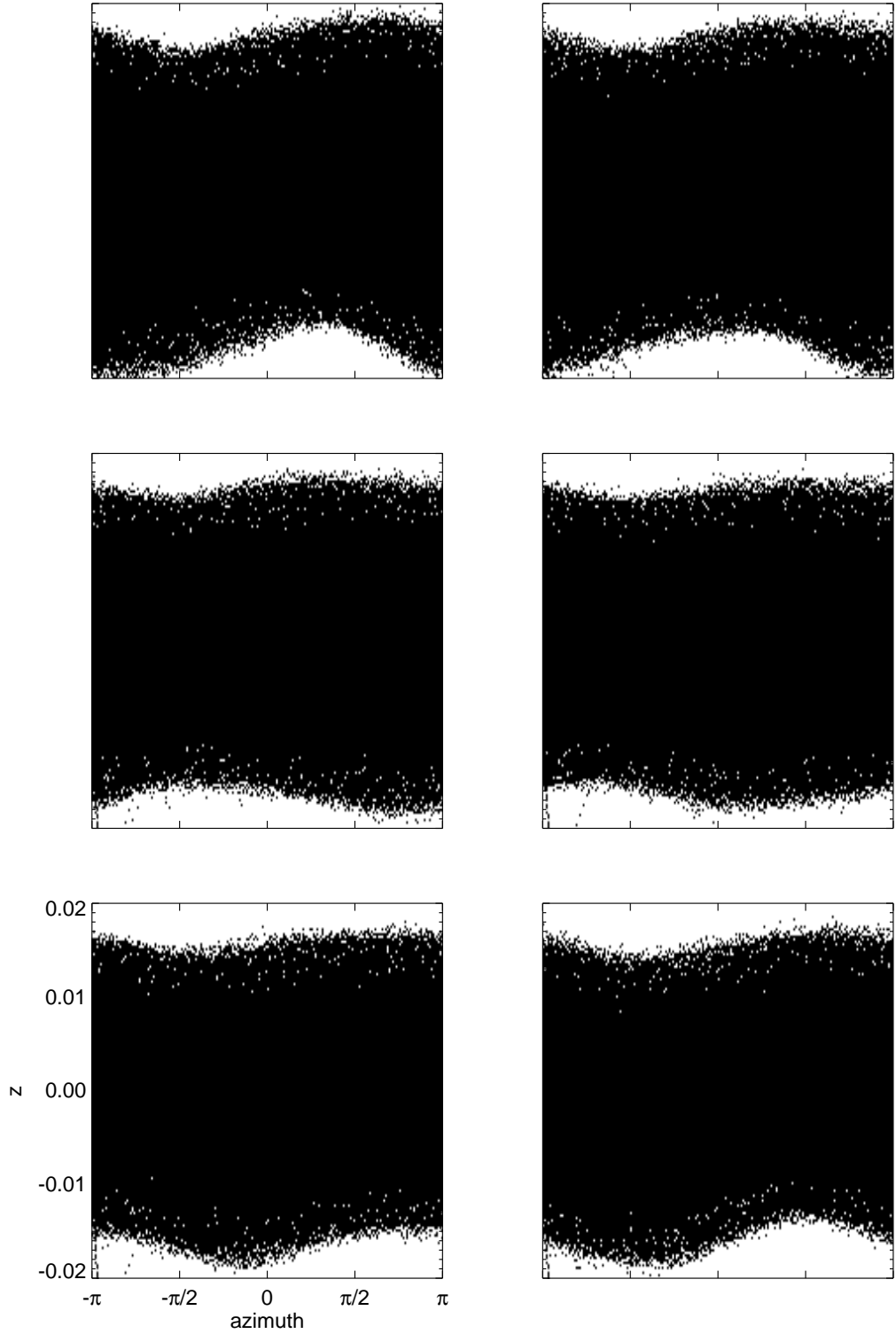,width=15cm}
\caption{Disc silhouette at times t=22,23,24,25,26 and 27. Panels 4
ans 5 show an almost flat disc and correspond to the minimum in warp
strength (see figure~\ref{discwarp}).
}
\label{sequence2}
\end{figure*}

When the magnetic field was switched on in the second calculation, the
disc immediately responded (solid line). Instead of increasing to a
steady value however, the warp mode strength oscillated with a period 
$P_{\rm w} = 0.987 \pm 0.003 \,P_{\rm orb}$ as the warp slowly 
precessed retrogradely 
in the inertial frame, and  the structure of the warp changed as
it moved relative to the binary. High time resolution animations of
the disc motion were also made
that confirmed this to be the precession period. Typical amplitudes of
$S^z$ for the magnetic calculation were of the order of $10^{-3}$,
about a factor ten or twenty less than for the tilted disc. These amplitudes
then agree with the vertical structure shown in figure~\ref{sequence2}.

There was a secular decline in warp amplitude. 
Now in these simulations we had $\alpha = 1$ and disc scale height
$H \simeq 0.04 r$. One can estimate the time scale for  viscous forces
to flatten a warp $t_{\rm f} \simeq \alpha (r/H)^2\,
\Omega^{-1}$. At the disc outer edge $t_{\rm f}\simeq 300$, not
entirely inconsistent with  the decay time scale apparent in
figure~\ref{discwarp}, particularly when one considers the disc
shrinkage over the course of the magnetic calculation. Reducing the
effective $\alpha$ of the disc would result in more rapid damping of
the warp.

\begin{figure}
\psfig{file=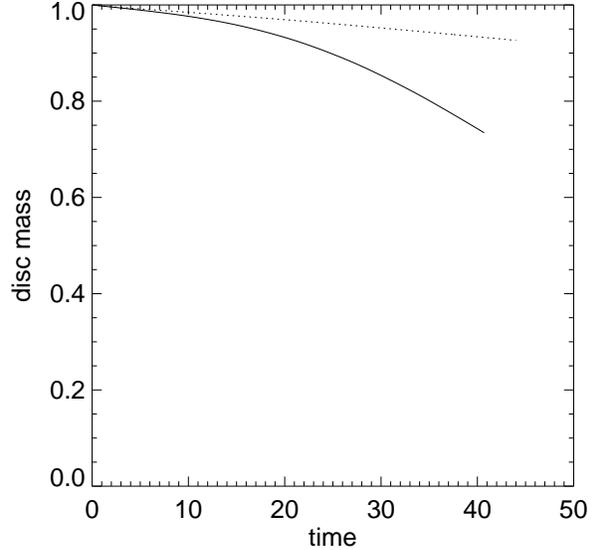,width=8cm}
\caption{Disc mass for the magnetised (solid line) and unmagnetised
(dotted line) simulations.}
\label{discmass}
\end{figure}

\begin{figure}
\psfig{file=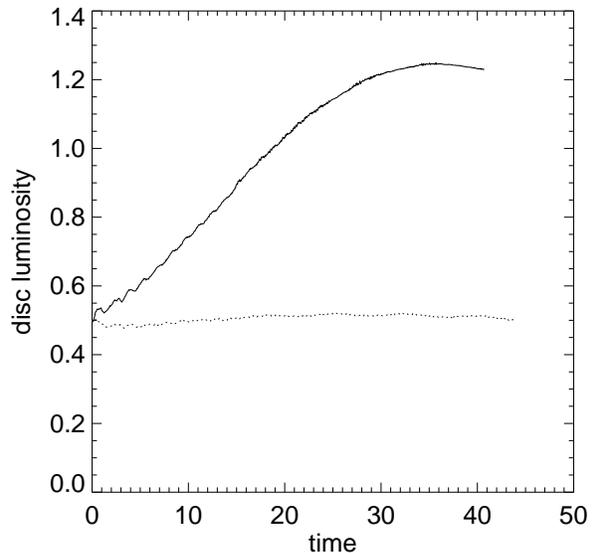,width=8cm}
\caption{Disc dissipation for the magnetised (solid line) and unmagnetised
(dotted line) simulations.}
\label{discdiss}
\end{figure}

The applied magnetic torque swiftly removed angular momentum from the
outer disc, causing it to shrink. Figure~\ref{discrad} shows the evolution of
the average disc radius ($R_{\rm d}$), which we define to be the
radius of a circle with the same area as the disc. The figure
clearly shows that the disc shrinks after the application of the 
magnetic field. This was predicted by the earlier treatment of
Pearson, Wynn and King (1997) and is a consequence of the 
relation $t_{\rm visc}(R_{\rm d}) \sim t_{\rm mag}(R_{\rm d})$. 
Figure~\ref{discmass} shows the
mass declined more rapidly for the torqued disc than for the reference
calculation. Energy dissipation
(which can be equated to bolometric
luminosity for the disc)
due to viscosity in the two discs 
is plotted in figure~\ref{discdiss}. The
sharp rise in dissipation  correlates with an accelerated rate of
accretion onto the white dwarf. It is
no surprise therefore that the warp strength declined in 
figure~\ref{discwarp}.

\begin{figure*}
\psfig{file=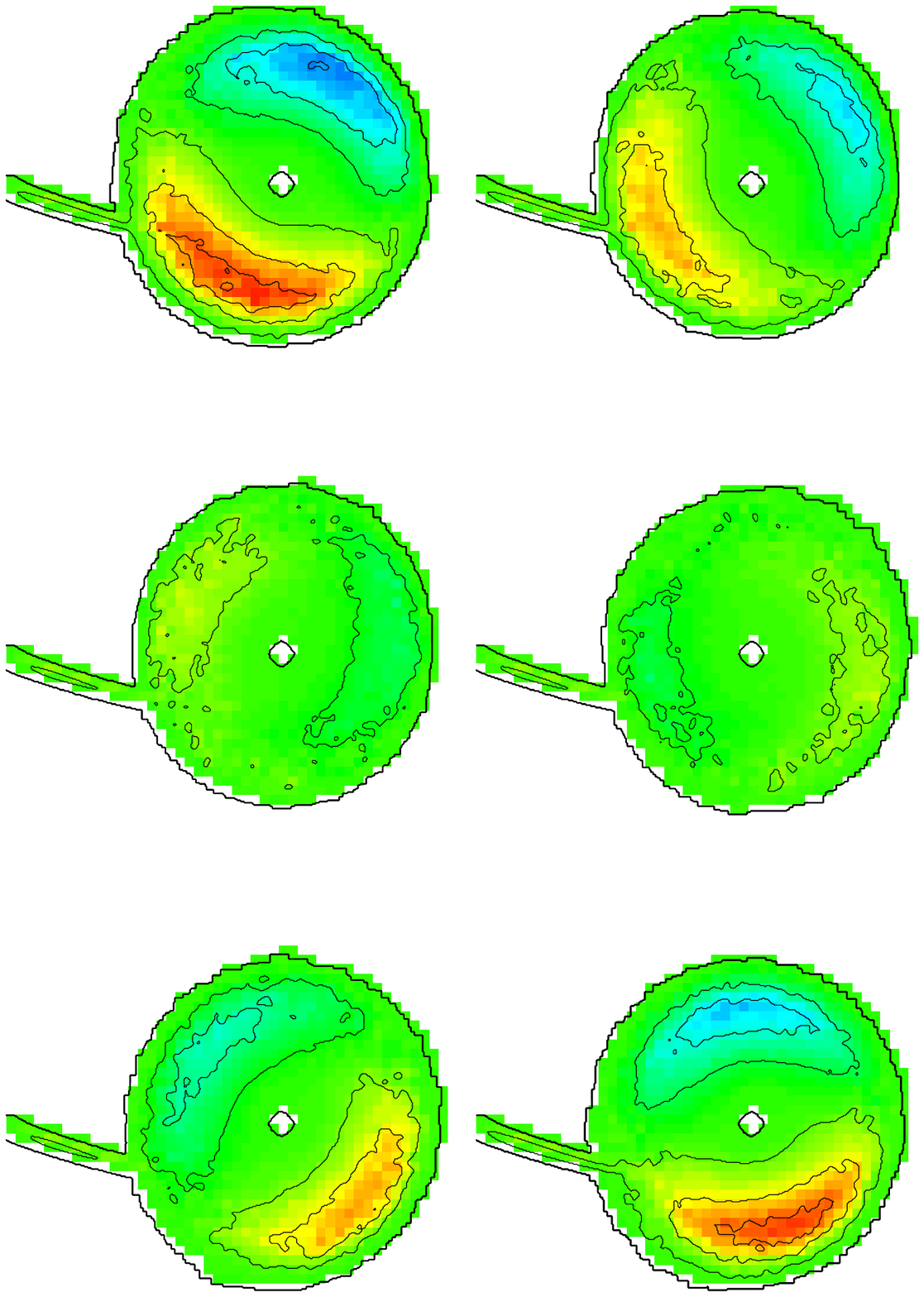,width=15cm}
\caption{Vertical structure of the disc at times t=22,23,24,25,26 and
27 (from top left to bottom right).  
This figure corresponds to figure~\ref{sequence2}.
}
\label{height2}
\end{figure*}

Figures~\ref{sequence2}
and~\ref{height2} show the changes in silhouette and plan view
respectively over the time period $t=22 - 27$. The retrograde precession of the warp
can clearly be seen in figure~\ref{height2}. The
variations in warp amplitude are also apparent. The warp almost
disappeared when it was
 $180^\circ$ out of phase with the magnetic
torque (in between panels 3 and 4). Even then however the disc was not
flat, but the vertical structure that was present at that time was symmetric about the
midplane (figure~\ref{sequence2} panels 3 and 4).
Half a binary period later (between panels 6 and 1 as the structure
changed cyclically) 
the warp was reinforced by the
magnetic torque and the disc was most distorted. In panel 1 for example
the scale height
varied by as much as 30 \% around the circumference of the disc. Such variations are
sufficiently large to be observable in nearly edge on systems. Again
we refer to the Billington et al. (1996) analysis of ultraviolet
superdips in OY Car. 

A simple toy model, constructed by the anonymous referee, is very
helpful in understanding the results shown in figures~\ref{height1},
\ref{discwarp} and \ref{height2}. Consider a single annulus of the
disc, and neglect pressure and viscosity. If the inclinations are
small, then the tilt of the annulus can be described by a variable
${\bf W}(t) = l_{\rm x} + i\,l_{\rm y}$ (where $x$ and $y$ are spatial
coordinates measured in the binary frame). 
${\bf W}(t)$ obeys a linear equation of the form
\begin{equation}
\frac{d{\bf W}}{dt}= i \omega_{\rm p}\,{\bf W} - \gamma\,({\bf W}-{\bf W_{\rm B}}),
\end{equation}
where $\omega_{\rm p} < 0$ is the tidal precession rate in the binary
frame, and $\gamma$ represents the effect of the magnetic drag. In the
absence of precession, the drag forces would bring  the disc tilt
${\bf W}$ to coincide with ${\bf W_{\rm B}}$ 
(determined by the orientation of the
secondary magnetic field) on a timescale $1/\gamma$. Starting with
zero initial tilt, we have
\begin{equation}
{\bf W}(t) = {\bf W_{\rm B}} \,\frac{\gamma}{\gamma - i\,\omega_{\rm p}}\,
[1-\exp{(i\,\omega_{\rm p} - \gamma)t}].
\label{eq:tilt}
\end{equation}
The oscillations in tilt amplitude in figure~\ref{discwarp} are neatly
explained as $|{\bf W}|$ is oscillatory with frequency $\omega_{\rm
p}$. Equation~\ref{eq:tilt} also shows that the warp's precession is a
transient response to the initial kick of the magnetic field. The
steady state disc response will be a disc warp that is stationary in 
the binary frame.

Negative superhumps have not been subject to the same degree of
scrutiny as their positive counterparts.
We are confident that positive superhumps arise in a precessing disc
(see most recently Rolfe, Haswell \& Patterson, 2001). Observations of
negative superhumps are limited to photometry, with no information
regarding either temperature or location of the emitting region.
Patterson et al. (2001) describes a persistent negative superhump 
observed in the nova-like variable V751 Cygni, and states 
that ``... many nova-like variables in their bright state and with
$P_{\rm orb} = 3-4$ hr show negative superhumps''. They do not show
the close association with superoutbursts that normal superhumps do,
and they do not appear in very low mass ratio systems (i.e. those
below the period gap). 

Positive superhumps are caused by a tidal resonance which can only act
upon an accretion disc if the ratio of  secondary to primary
stellar masses $ q \ale 1/3$ (Lubow 1991; Murray, Warner \&
Wickramasinghe 2000). Lubow (1992) showed that a second  resonance
that acts to tilt the disc, almost exactly overlays the eccentric
resonance. The tilt resonance is however much weaker, and
simulations described in Murray \& Armitage (1998) failed to produce a
disc warp, but with 30 000 particles in the calculations any detection
of a warp would have been marginal.  

It is worth revisiting the problem, making
use of the much higher resolution now available to us ($10^6$
particles) and also considering a higher
mass ratio. The strength of the tilt resonance goes as $q^2$ and mass
ratios of observed negative superhumpers are closer to $1/3$ rather
than the $3/17$ used in Murray \& Armitage. The calculations presented
in this paper give us confidence that smoothed particle hydrodynamics
simulations are an appropriate tool for studying the evolution of warp
modes in discs, however the warps are generated.

\section{Conclusions}
We have followed the evolution of a warp in a close binary accretion
disc. An initially planar disc was subjected to a force  that modelled
the effects of an inclined magnetic dipole centred on the mass donor
star. The magnetic field was assumed to be fixed in the binary frame.

Other authors have modelled the evolution of discs with pre-existing
warps, or of discs that were misaligned with the binary plane. However
these are the first hydroydnamic simulations to follow the initial
development of a warp, and the first to show how the warp structure
varies as it precesses.

The disc's initial response to the inclined field was complicated by
spiral shocks excited by the secondary's tidal field. However the
spiral shocks are strongly damped by the developing warp. In turn the warp
structure rapidly simplified and steady, uniform retrograde precession (in the
inertial frame) occurred.

The amplitude and structure of the warp depended upon 
its phase with respect to the
perturbing force (and hence of the binary). 
The warp was maximum when it was reinforced by the
magnetic force. However $180~\deg$ later, when the warp and the force
were  anti-phased, the disc was almost flat. We measured the period of
the motion of the warp with respect to the binary to be $P_{\rm w}=0.987 \pm
0.003 \,P_{\rm orb}$. 

We suggest therefore that negative superhumps could be explained not
simply via the retrograde precession of a warped disc, but in terms of
changes in the discs vertical structure that occur on the same
period, $P_{\rm w}$. Because of the very limited observations of
negative superhumps we can but hypothesise as to the
precise location of the luminosity variations. One possibility is the
changing disc aspect presented to the accretion stream as the warp
precesses. Alternatively, negative superhumps may be the result of
radiation from the central source being reprocessed over a changing
disc surface.

In subsequent papers already in preparation 
we shall discuss both the evolution to steady state of discs in
magnetised systems, and the influence of magnetised accretors on disc structure.
\section*{acknowledgements}
Calculations for this paper were performed on the UK Astrophysical
Fluids Facility SGI Origin 3000, and on the Leicester Mathematical
Modelling Centre Origin 2000. The latter machine (HEX) was purchased with an
EPSRC grant. Phil Armitage, Jim Pringle and Andrew King provided
ideas and influenced the content. Chakrabarty is employed on a
postdoctoral fellowship funded by a grant from the Leverhulme Trust.
The code was parallelised by
Michael Truss. Finally, many thanks to the anonymous referee who cast
a fresh eye over our work and presented a simple explanation for the
simulation results.


\begin{thebibliography}{}
\bibitem{Applegate}
Applegate J.H., and Patterson, J., 1987, ApJ, 322, L99
\bibitem{Bianchini}
Bianchini, A., 1990, AJ, 99, 1941
\bibitem{Billington}
Billington,I. et al., 1996, MNRAS, 279, 1274
\bibitem{Howell}
Howell, S.B., Ciardi D.R., Dhillon V.S., and Skidmore, W.,
2000, ApJ, 530, 904
\bibitem{King}
King, A.R., 1993, MNRAS, 261, 144
\bibitem{KFW}
King, A.R., Whitehurst, R., Frank, J., 1990, MNRAS, 244, 731
\bibitem{Kraft}
Kraft, R.P., 1967, ApJ, 150, 551
\bibitem{DongLai}
Lai, D. 1999, ApJ, 524,1030
\bibitem{LNPT}
Larwood. J.D., Nelson, R.P., Papaloizou, J.C.B., Terquem, C. 
1996, MNRAS, 282, 597
\bibitem{Lub1}
Lubow, S.H., 1991, ApJ, 381, 259
\bibitem{Lubow}
Lubow, S.H., Shu, F.H., 1975, ApJ, 198, 383 
\bibitem{Mestel}
Mestel L., and Spruit, H., 1987, MNRAS, 266, 57
\bibitem{JRM1}
Murray, J.R., 1996, MNRAS, 279, 402
\bibitem{Muzz3}
Murray, J.R., 1997, 
Accretion Phenomena and Related Outflows; IAU Colloquium 163. 
ASP Conference Series; Vol. 121; ed. D. T. Wickramasinghe; G. V.
Bicknell; and L. Ferrario, p.770
\bibitem{JRM2}
Murray, J.R., 1998, MNRAS, 297, 323
\bibitem{JandP}
Murray, J.R., Armitage, P.J., 1998, 300, 561
\bibitem{MartinJianke}
Murray, J.R., de~Kool, M., Li, J., 1999, ApJ, 515, 738 
\bibitem{BanD}
Murray, J.R., Warner, B., Wickramasinghe, D.T., 2000, MNRAS, 315, 707 
\bibitem{NelsonPap}
Nelson, R.P., Papaloizou, J.C.B., 1999, MNRAS, 309, 929
\bibitem{NelsonPap2}
Nelson, R.P., Papaloizou, J.C.B., 2000, MNRAS, 315, 570
\bibitem {O'Donoghue}
O'Donoghue, D., New Astronomy Reviews, Volume 44, Issue 1-2, p 45-50
\bibitem{Gordon}
Ogilvie, G.I. \& Dubus, G., 2001, MNRAS, 320, 485
\bibitem{PattersonK}
Patterson, J., 1999, 61, in Disk Instabilities in Close Binary Systems, ed.s 
S. Mineshige \& J.C. Wheeler, Universal Academy Press
\bibitem{Patterson2}
Patterson, J., et al., 2001, PASP, 113, 72
\bibitem{Pearson}
Pearson, K.J., Wynn, G.A. and King A.R., 1997, MNRAS, 288, 421 
\bibitem{Pringle}
Pringle, J.E., 1996, MNRAS, 281, 357
\bibitem{Rolfe}
Rolfe, D.J., Haswell, C.A., Patterson, J., 2001, MNRAS, 324, 529
\bibitem{Skumanich}
Skumanich, A., 1972, Apj, 171, 565
\bibitem{TP}
Terquem, C., Papaloizou, J.C.B., 2000, A\&A, 360, 1031
\bibitem{Michael}
Truss, M.R., Murray, J.R., Wynn, G.A., Edgar, R.G., 2000, MNRAS, 319, 467 
\bibitem{Verbunt}
Verbunt, F., and Zwaan, C., 1981, A\&A, 100, L7
\bibitem{Warnerbook}
Warner, B., 1995b, Cataclysmic Variable Stars. Cambridge University
Press, Cambridge
\bibitem{wdwarp}
Wynn, G.A., Murray, J.R., Chakrabarty, Dalia, 2001, in preparation
\bibitem{MontWoodSimp}
Wood, M.A, Montgomery, M.M., Simpson, J.C., 2000, ApJ, 535, L39
\bibitem{Boffin}
Yukawa,H., Boffin, H.M.J., Matsuda, T., 1997, MNRAS, 292, 321
\end{thebibliography}
\end{document}